\documentstyle[10pt]{article}
\input psfig.tex
\begin{document}
\vfill
\begin{center}{\large\bf New Results on Perturbative Color Transparency

 in
Quasi-Exclusive Electroproduction}\end{center}\par
\vskip 0.5cm
\centerline{Pankaj Jain$^a$, Bijoy Kundu$^a$, John Ralston$^b$ and
Jim Samuelsson$^a$}
\vskip 0.5cm
\centerline{$^a$Physics Department, I.I.T. Kanpur, 208016, India}
\centerline{$^b$Department of Physics \& Astronomy, University of Kansas,}
\centerline{
Lawrence, KS 66045, USA }\par
\vskip 1.0cm
PACS number(s): 13.40.Fn, 12.38.Bx, 14.20.Dh
\vskip 2.0 cm
%\baselineskip=2\baselineskip
\centerline{\bf Abstract} We review the perturbative QCD
formalism of hadronic electromagnetic form factors and the color
transparency ratio for quasi-exclusive electroproduction of the proton and
pion from nuclear targets. We have completed the first full calculations
including all leading order quark subprocesses and integrations over
distribution amplitudes, including Sudakov effects. For the case of the
proton, the calculated result shows scaling beyond $Q^2=10$ GeV$^2$. The
calculation incorporating filtering due to the nuclear medium is cleaner
than the corresponding calculation in free space because of attenuation of
large distance amplitudes. We find that the color transparency ratio is
rather insensitive to theoretical uncertainties inherent in the
perturbative formalism, such as the choice of the hadron distribution
amplitude.

\bigskip

\section{Introduction}

The practical applicability of perturbative QCD to exclusive processes
\cite{GRF,GPL,CZ} such as hadronic electromagnetic form factors cannot yet
considered   to be settled. It has been argued that even at the highest
momenta explored so far in the laboratory, the dominant contribution to
form factors comes from the end point regions of the wave function,
where the perturbative treatment fails \cite{Isg,Rad}.
In the case of hadron-hadron scattering there exist further difficulties,
such as the common failure of the helicity conservation selection rules
to agree with experimental data.
However in nearly all experiments one finds that
the naive prediction for quark counting
scaling laws tend to agree very well with data. In view of the
problems listed, this is quite mysterious, since so far there does
not exist any alternate mechanism which can explain these scaling laws.

An interesting prediction of perturbative
QCD is color transparency \cite{BrodMuel}.
At large momentum transfers only the short distance components of the
hadron wave function can contribute to exclusive processes. Since
the total cross section of hadrons $\sigma$ is inversely proportional
to their area $b^2$, the strong interactions of these
hadrons is expected to be reduced. If we consider quasi-exclusive
electron-nucleus  scattering,
$eA\rightarrow e'p(A-1)$, where $A$ is nuclear number,
then the nucleus is predicted to be transparent to all protons participating
in this process.  This is an asymptotic argument applicable for fixed $A$
as $Q^2 \rightarrow\infty$.  Experimentally, however, we can only take
the limit of large $A$ and moderately large $Q^2$, in which such processes
appear to be more complicated and much more interesting.  One picture that
is emerging \cite{JPR}
is that exclusive processes in free space get significant
contribution from perturbatively calculable hard amplitudes
but also have non-negligible soft contamination.
The corresponding nuclear processes, however, may be much cleaner
\cite{JB,BT88} because the large quark separations will be
strongly attenuated in nuclear medium. This
phenomenon, called nuclear filtering, has some experimental support.
Experimentally one finds that the fixed-angle free space process
$pp'\rightarrow p''p'''$ \cite{Car} shows significant oscillations at 90
degrees as a function of energy.   These oscillations are not a small
effect, but roughly 50\% of the $1/s^{10}$ behavior, and are interpreted as
coming from interference of long and short distance amplitudes. The
corresponding process in a nuclear environment $pA\rightarrow p'p''(A-1)$
shows no oscillations, and obeys the pQCD scaling
power law better than the free-space data \cite{JB,BT88}. The $A$
dependence, when analyzed at fixed $Q^2$, shows statistically significant
evidence of reduced attenuation \cite{JainRalPRD}.

\section{Formalism}

We  briefly review the framework for calculation of hadronic
form factors following Li and Sterman \cite{LS}.  It has long been known
that the transverse separation of quarks in free space
reactions is controlled by effects known as the Sudakov form factor.  The
pion form factor is the simplest example. Li and Sterman  included Sudakov
effects here, arguing that a perturbative treatment becomes fairly reliable
at momenta of the order of 5 GeV.  As low as 2 GeV, it was found that less
than 50 \% of the contribution comes from the soft region.

Let $b_{ij}$ be the transverse separation between quarks $i$ and $j$, or
$b$ the corresponding quantity for a single pair of quarks.  An essential
feature is the inclusion of $exp(-S)$, a Sudakov form factor which
suppresses the large $b$ region. Including the $b$ dependence, the pion
electromagnetic form factor can be written as, \begin{eqnarray} F_\pi(Q^2)
= \int dx_1
dx_2{d\vec b\over (2\pi)^2} {\cal P}(x_2,\vec b,P',\mu) T_H(x_1,x_2,\vec
b,Q,\mu){\cal P}(x_2,\vec b,P,\mu)\ . \label{fpi} \end{eqnarray} where $${\cal
P}(x,b,P,\mu) = exp(-S)\times \phi(x,1/b) + O(\alpha_s(1/b))\ ,$$ plays the
role
of the hadron wave function, $\phi(x,1/b)$ is the meson distribution amplitude,
$P$ and $P'$ are the incident and outgoing pion momenta respectively, and
$S$ is
the Sudakov form factor. The improved factorization used in \cite{LS} retains
the intrinsic transverse momentum $k_T$ dependence in the gluon propagator,
since $k_T$ need not be small compared to $\sqrt{x_1x_2}Q$, if one of the $x_i$
get close to zero. The variable $b$ in Eq. \ref{fpi} is conjugate to $k_{T1} -
k_{T2}$, where $k_{T1}$ and $k_{T2}$ are the transverse momenta of the incident
and outgoing pions. As long as $x_1$ and $x_2$ are not close to their
endpoints,
the dominant scale in the scattering is $\sqrt{x_1x_2}Q$ and the small $b$
region dominates the amplitude. Close to the end points of $x_1$ or $x_2$,
$\sqrt{x_1x_2}Q$ may become very small. However, the dominant scale in this
region is $1/b$, which is again not too small since the large $b$ region is
strongly damped by the Sudakov form factor. The results for the free space
form
factor for pion using this procedure are given in \cite{LS}. The authors show
that at $Q^2=5$ GeV$^2$, something like 90\% of the contribution comes from a
region where $\alpha_s/\pi$ is less than 0.7 and hence could be regarded as
perturbative.

\medskip

The nuclear medium modifies the quark wave function such that
\cite{RP90,stmalo}
\begin{eqnarray} {\cal P}_A(x,b,P,\mu) = f_A(Q^2,b){\cal P}(x,b,P,\mu),
\end{eqnarray} where ${\cal P}_A$ is the wave function inside the medium and
$f_A$ is the nuclear filtering amplitude. We use a simple model for $f_A$,
$$f_A = exp(-\int dz\ \sigma \rho)\ \ .$$
The effective inelastic cross section $\sigma$ is known to scale like $b^2$ in
QCD, where $b$ is the size of the hadron. We parametrize it as $k b^2$ and
adjust the value of $k$ to find a reasonable fit to the experimental data.

The situation for the proton form factor \cite{LS1}
is somewhat more complicated than that of the pion; we do not have the
space for all details here which are given in Ref. \cite{BL,Bijoy99}. 
There has been some controversy regarding the
proper choice of the
infrared cutoff in the Sudakov exponent. In the case of pion this was
simply the quark-antiquark separation $b$. The choice proposed in
\cite{BK} uses the largest distance between the three quarks as the cutoff. It
was found that this gave results about 50 \% smaller than experiments.
Perhaps this is the right direction, if indeed other wave functions (and in
particular, non-zero quark angular momentum) contribute heavily in free space.
On the other hand, in \cite{BL} it was observed that the largest distance does
not correspond to a physical size of the three quark system. A more appropriate
choice might be obtained by considering the triplet of valence quarks
as a quark-diquark system.
This choice takes the maximum value of the distance between quark and
diquark as
the effective cutoff in the Sudakov exponent. This essentially amounts to
using a scale $cw$ for infrared cutoff,
with $c\approx 1.14$, where $w$ is the inverse of the
largest distance between any two valence quarks in proton.
 Remarkably, this small
modification leads to results in good agreement with the experiment \cite{BL}.

From investigations of the proton form factor in free space, it seems that
Sudakov effects eliminate about 50 \% of the contribution from the soft
region. The Sudakov filtering in free space does something useful, but does
not seem to be sufficient to make present free-space calculations totally
reliable. The same diagrams for Sudakov effects of course occur in a
nuclear environment. In addition, there are much stronger interactions with
the nuclear target, when one goes from pure ``vacuum filtering" by Sudakov
to nuclear filtering.  We find that nuclear medium
eliminates much of the remaining 50 \% of the soft region. These are the
first full calculations of these ideas within perturbative QCD.  We find
that the main uncertainty in the nuclear calculation arises from
uncertainties in nuclear medium itself, in particular, in uncertainties in
the nuclear spectral
functions and correlations. With standard assumptions one can proceed with
the calculation essentially using zero parameters and no model dependence.
However, we find that numerical differences between models of nuclear
matter are large enough to cause significant uncertainties. Indeed,
comparison with data shows
that the uncertainties in the nuclear spectral functions and the nuclear
correlations now dominate the theoretical uncertainties, and are larger
effects than, for example, the dependence on the hadron distribution
amplitude.

\begin{figure} [t,b] \hbox{\hspace{6em}
\hbox{\psfig{figure=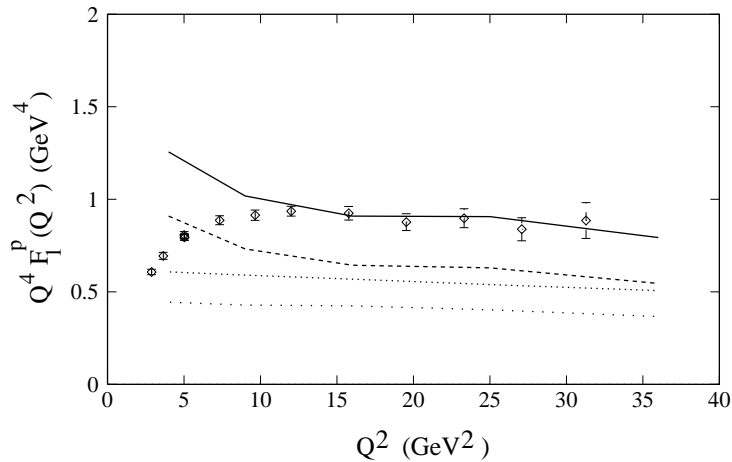,height=6cm}}} \caption{The $Q^2$ dependence of
the proton form factor $Q^4F_1$ using the KS wave function ($c=1.14$, solid
line;
$c=1$, dense-dot line) and for the CZ wave function ($c=1.14$, dashed line;
$c=1$, dotted
line). The experimental data are also shown.}
\label{fig1} \end{figure}

\begin{figure} [t,b] \hbox{\hspace{6em}
\hbox{\psfig{figure=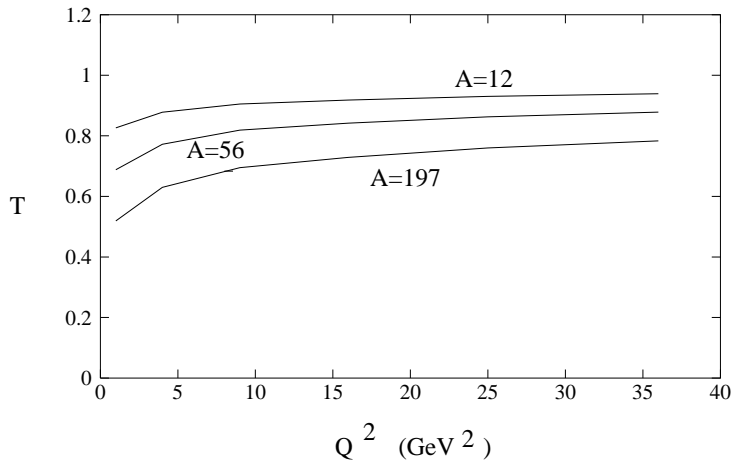,height=6cm}}} \caption{The calculated pion
transparency ratio for different nuclei.} \label{fig2} \end{figure}

\section {Results and Discussions}

The results for free space proton form factor are
shown in Fig. \ref{fig1}. An important  feature of this result,
which is independent of the details of the wave function, is that it shows
scaling for $Q^2$ larger than about 10 GeV$^2$. This is a nontrivial
confirmation that $Q^2$ indeed dominates over the intrinsic momentum
$k_T^2$.

\begin{figure} [t,b] \hbox{\hspace{6em}
\hbox{\psfig{figure=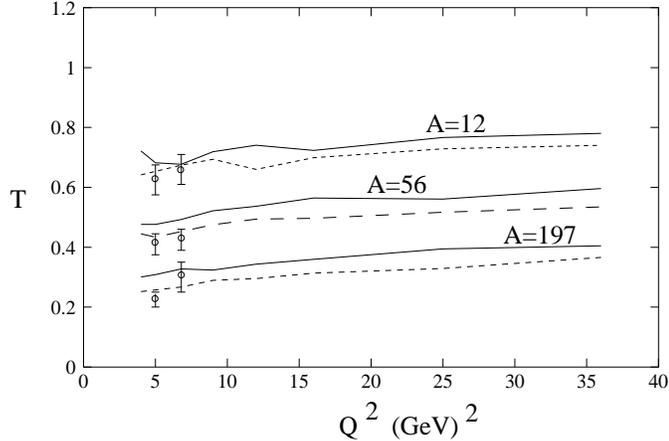,height=6cm}}} \caption{The calculated
transparency ratio for the proton for different nuclei.
The experimental points are
taken from Ref. [19,20]. The solid curves are calculated with
$k=5$ and the dashed
curves with $k=6$.} \label{fig3}
\end{figure}

In Fig. \ref{fig2}, we  show results for color transparency for
electroproduction of pions for different nuclei using the CZ wave function.
Here we adjust the value of k, corresponding to the pion attenuation
cross-section of 25-30 mb for a pion size of about 0.8 fm. The predicted
results are shown for k=4. The precise value of $k$ might best obtained by
making a fit to the data for color transparency after it becomes available,
or perhaps by detailed comparison with diffractive calculations.  Compared
to the asymptotic wave functions, the results
for $T$ change by less than 3\% for $Q^2$ larger than 10 GeV$^2$.

\begin{figure} [t,b] \hbox{\hspace{6em}
\hbox{\psfig{figure=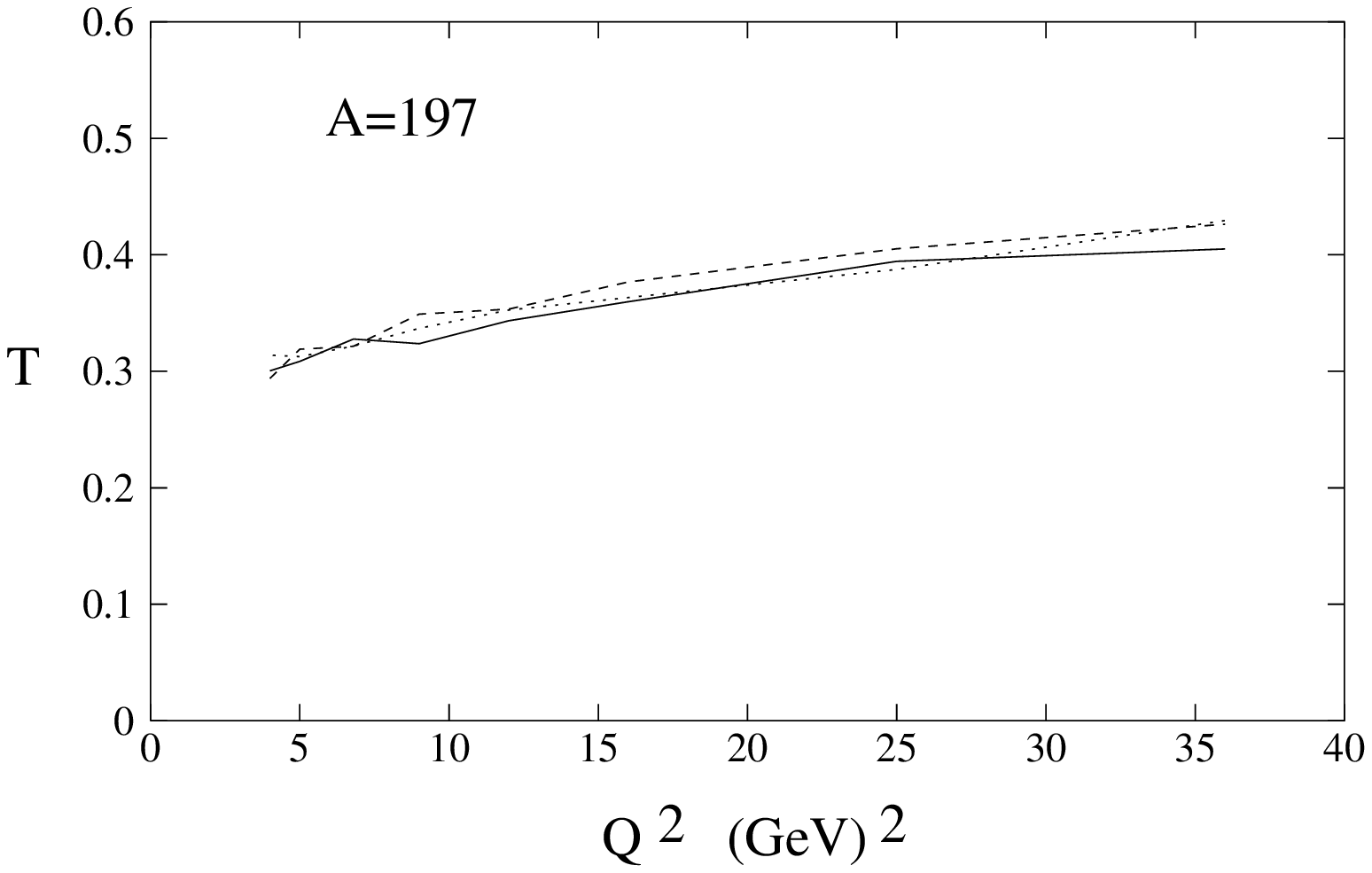,height=6cm}}} \caption{ The sensitivity of the
calculated transparency ratio to different proton wave functions.  Slight
oscillations are an artifact of the Monte Carlo integrations.  The solid curve
is calculated with the KS wave function, as in Fig. 3, and the dotted curve is
calculated with the CZ wave function; both curves use the infrared cutoff
parameter $c=1.14$. For reference,
the dashed curve shows the
result for the cutoff proposed in Ref. 18, which
amounts to setting $c=1.0$, using the KS wave function.
The calculations are shown for $A=197$.}
\label{fig4} \end{figure}

The results for the proton transparency ratio are given in Fig. \ref{fig3}.
The parameter $k$ in the attenuation cross section $\sigma=kb^2$
was chosen so as to provide a reasonable fit to the experimental data
\cite{Mak,Neill}. We find that a value of $k=6$ gives a reasonable fit.
Taking the attenuation cross section of normal protons to be 36 mb, this
corresponds to a typical $b$ of about 0.77 fm, which is a reasonable
estimate of the proton size. Since the data for $T$ is available only in
the region where the calculated free space form factor is in disagreement
with the experimental result, the value of $k$ obtained by this procedure
cannot be taken too seriously. In fact, parameter $k$ would be best
obtained by fitting to the experimental value of $T$ after it is measured
at higher energies. A reasonable range of $k$ values, which we take to be
$k=5$ and $k=6$, corresponds to $b$ values of 0.85 fm and 0.77 fm
respectively, and has been used in the figure.

We have also checked the dependence of our result on the infrared cutoff
parameter $c$ and the choice of the wave function. We find in Fig.
\ref{fig4} that the results for transparency ratio change very little if we
use the CZ wave function instead of the KS. This is a surprising result,
and one of the basis of our claim that the dominant uncertainty in
transparency ratio may be due to the nuclear model itself.

\section {Conclusion} We have reviewed the calculation of hadronic
electromagnetic form factors and color transparency using perturbative
QCD.  We find a slow rise in the transparency ratio for energies that can
be probed in the future at CEBAF and ELFE.  As discussed elsewhere
\cite{JainRalPRD,JPR} , precision experiments can discover color
transparency even with a slow rise in $Q^2$ by measuring the $A$ dependence
at fixed moderately large $Q^2$.   Due to filtering of long distance
components in the medium, the nuclear calculation is considerably cleaner
compared to the free space calculation. We also find rather remarkable
insensitivity of the transparency ratio to present theoretical
uncertainties in the perturbative QCD treatment, such as the choice of the
distribution amplitude.  To further improve the accuracy of predictions for
color transparency ratio, it is necessary to improve the modelling of
nuclear medium which now appears to be the dominant source of error.

\section * {Acknowledgements}
We thank Hsiang-nan Li for many useful discussions.  Financial support for
this work was provided by the Board
of Research in Nuclear Sciences (BRNS),
the Crafoord Foundation and the DOE grant 85ER401214.

 \end{document}